\documentclass[prc,twocolumn,showpacs,showkeys,preprintnumbers,amsmath,amssymb]{revtex4}

\usepackage{graphicx}
\usepackage{dcolumn}
\usepackage{bm}

\begin{document}

\title{How does breakup influence the total fusion of
$^{6,7}$Li at the Coulomb barrier?}

\author{A. Diaz-Torres\footnote{Email: A.Diaz-Torres@theo.physik.uni-giessen.de}}
\affiliation{Institut f\"ur Theoretische Physik der
Justus--Liebig--Universit\"at Giessen, Heinrich--Buff--Ring 16,
D--35392 Giessen, Germany}

\author{I.J. Thompson}
\affiliation{Physics Department, University of Surrey,
Guildford GU2 7XH, United Kingdom}

\author{C. Beck}
\affiliation{Institut de Recherches Subatomiques, UMR
75000,CNRS--IN2P3 et Universit\'e Louis Pasteur, 23 rue du Loess,
B.P. 28,F--67037 Strasbourg, Cedex 2, France}


\begin{abstract}

Total (complete + incomplete) fusion excitation functions of
$^{6,7}$Li on $^{59}$Co and $^{209}$Bi targets around the Coulomb
barrier are obtained using a new continuum discretized coupled
channel (CDCC) method of calculating fusion. The relative
importance of breakup and bound-state structure effects on total
fusion is particularly investigated. The effect of breakup on
fusion can be observed in the total fusion excitation function.
The breakup enhances the total fusion at energies just around the
barrier, whereas it hardly affects the total fusion at energies
well above the barrier. The difference between the experimental
total fusion cross sections for $^{6,7}$Li on $^{59}$Co is notably
caused by breakup, but this is not the case for the $^{209}$Bi
target.

\end{abstract}



\maketitle

\section{Introduction}

The effect of breakup of weakly bound projectiles on fusion has
been extensively investigated in recent years both experimentally
\cite{Yoshida,Takahashi,Kolata,Rehm,Dasgupta,Aguilera,Trota,Moraes,
Mukherjee,Signorini,Navin,Padron,Dasgupta_new,Beck_new} and
theoretically
\cite{Hussein2,Hagino,Alexis1,Alexis2,Keeley1,Keeley2,Yabana,Alamanos},
but there is not yet a definitive conclusion. Experimental works
discuss the effect of breakup on fusion by comparing experimental
fusion excitation functions to either realistic theoretical
predictions which do not include couplings to the breakup
channels, e.g. \cite{Dasgupta,Trota,Navin,Dasgupta_new}, or to
experimental fusion excitation functions of well-bound (reference)
projectiles for which the breakup is expected to be weak, e.g.
\cite{Rehm,Moraes,Signorini,Padron}.

In fusion of weakly bound nuclei, two independent fusion processes
can be distinguished, namely complete fusion and incomplete or
partial fusion. The total fusion is the sum of these processes
(complete + incomplete). These two types of fusion processes are
connected to the dynamics of the projectile fragments. A clear
definition of complete and incomplete fusion is necessary to
compare theoretical predictions to experimental data.
Theoreticians, e.g. \cite{Hussein2,Hagino}, and experimentalists,
e.g. \cite{Dasgupta,Dasgupta_new2}, give different definitions in
the literature. From a theoretical point of view we think that,
strictly speaking, complete fusion refers to the capture of all
the projectile fragments (from bound and breakup states) by the
target, whereas the incomplete fusion is related to the capture of
only some of those fragments. Experimentalists
\cite{Dasgupta,Dasgupta_new2} tend to define complete and
incomplete fusion as absorption of all the charge of the
projectile and of a part of that charge, respectively. These
definitions are only equivalent to the strictly theoretical ones
if all the projectile fragments are charged (e.g.,
$^{6}$Li=$\alpha $+$d,$ or $^{7}$Li=$\alpha $+$t$), but this is
not the case for projectiles like $^{9}$Be$=\alpha + \alpha + n$
or $^{11}$Be$=^{10}$Be$ + n$. It is important to note that in
fusion experiments with $^{9}$Be \cite{Dasgupta,Signorini} and
$^{11}$Be \cite{Signorini} it is still unclear what happens to the
valence neutron after the reaction. If we follow the experimental
definitions, the total and complete fusion would be the same for
the reaction $^{11}$Be + $^{209}$Bi \cite{Signorini}, since only
the capture of the stable $^{10}$Be core has been observed so far.

Most experiments \cite{Yoshida,Takahashi,Kolata,Rehm,Aguilera,
Trota,Moraes,Mukherjee,Signorini,Padron,Beck_new} have only
measured the total fusion, whereas in
\cite{Dasgupta,Navin,Dasgupta_new} the complete fusion was
distinguished from the incomplete one. The importance of
distinguishing between complete and incomplete fusion in order to
observe complete fusion suppression above the barrier for
$^{6,7}$Li + $^{209}$Bi was pointed out in \cite{Dasgupta_new}.

Since the calculation of the complete and the incomplete fusion
cross section following either the strictly theoretical
definitions or the experimental ones is extremely complex,
simplifying models have been used up to now
\cite{Dasgupta_new,Hussein2,Hagino,Alexis1,Alexis2}. Using full
coupled channels calculations, complete fusion was interpreted in
\cite{Hagino,Alexis1} as absorption from projectile bound states,
and as incomplete fusion that from unbound states. In our approach
in \cite{Alexis1}, the total fusion cross section for $^{11}$Be +
$^{208}$Pb was unambiguously (referred to the strictly theoretical
definition) calculated, but this was not the case for the complete
and incomplete fusion independently. The unambiguous prediction of
complete and incomplete fusion cross sections is still a challenge
for all current fusion models \cite{Dasgupta_new}.

Following Ref. \cite{Hagino}, we interpreted in \cite{Alexis1} the
total fusion of $^{11}$Be on $^{208}$Pb as the absorption of the
center of mass (c.m.) of the projectile from either its bound or
breakup states. Since the mass of the $^{10}$Be core is much
larger than the neutron mass, such an absorption could ensure that
at least the charged core is captured. However, we doubt that this
approach can be used to study the total fusion of the two-cluster
projectiles $^{6,7}$Li because the two charged fragments
($^{6}$Li=$\alpha $+$d,$ and $^{7}$Li=$\alpha $+$t$) have similar
masses. In this case, the capture of the c.m. of the projectile is
not necessarily connected to the capture of the charged fragments.
Therefore, we will use for the present reactions two optical
potentials for the nuclear interaction between the projectile
fragments and the target. At the same time, the short-ranged
imaginary fusion potential defined in the c.m. coordinate of the
projectile and used in \cite{Alexis1}, will be removed. The
imaginary part of those optical potentials will also be
short-ranged (inside the Coulomb barriers) to ensure that the
absorption is associated with fusion channels only.

The present work particularly aims at (i) investigating the
relative importance of breakup and bound-state structure effects
(i.e. ground-state reorientation couplings, coupling to the bound
excited state ($^{7}$Li) and bare potential) for the total fusion
of $^{6,7}$Li on $^{59}$Co and $^{209}$Bi targets around the
Coulomb barrier, (ii) clarifying whether or not the enhancement or
suppression effect of breakup is shown in total fusion, and (iii)
testing the model with recent experimental data
\cite{Dasgupta_new,Beck_new}. In Sect. II, the theoretical
formalism is presented, whereas the results and the discussion are
shown in Sect. III. We draw conclusions in Sect. IV.

\section{Theory}

Calculations of total fusion cross sections for $^{6,7}$Li on
$^{59}$Co and $^{209}$Bi are carried out using a three-body model
\cite{Alexis1} with a new CDCC \cite{CDCC} method of calculating
fusion, i.e. with short-range fusion potentials for each fragment
separately. Full coupled channels calculations are performed with
the code FRESCO \cite{FRESCO}. The set of coupled equations
\cite{Alexis1} for the projectile-target radial wave functions is
solved with the usual scattering boundary conditions
\cite{FRESCO}.

The total fusion cross section is calculated in terms of that
amount of flux which leaves the coupled channels set (total
absorption cross section) because of the short-ranged imaginary
parts $iW_{F}$ of the optical potentials between the projectile
fragments and the target. This guarantees that at least one of the
charged fragments of the projectile is captured. The same
Woods-Saxon potential $W_{F}$ with parameters $W_{0}=-50$ MeV,
$r_{0}=0.8$ fm, and $a=0.1$ fm is used for the imaginary part of
the two optical potentials. The results depend only very weakly on
the parameters of this potential, as long as it is well inside the
Coulomb barrier and strong enough for the mean-free path of the
projectile inside the barrier to be much smaller than the
dimensions of $W_{F}$. The fusion cross sections for $W_{0}=-50$
MeV are those for $W_{0}=-10$ MeV changed by $\sim 1\%$. The use
of a short-ranged imaginary potential is equivalent to the use of
an incoming boundary condition inside the barrier for each
fragment to study fusion \cite{Brown}.

In calculations of total fusion cross sections, we simultaneously
include (i) the breakup of the projectile caused by inelastic
excitations to different partial waves in the continuum
(non-resonant and resonant breakup), induced by the projectile
fragments-target interactions (Coulomb+nuclear), and (ii) all
continuum (bound-continuum and continuum-continuum) and
reorientation couplings. By breakup we mean the elastic
dissociation of $^{6,7}$Li into two fragments only, namely alpha +
deuteron for $^{6}$Li and alpha + triton for $^{7}$Li, and not
further breakup of the deuteron or triton. The reorientation
couplings refer to the couplings of the quadrupole term of the
projectile fragment-target potentials, among the projectile target
partial waves, for the projectile in its ground state. Since we
will first focus on effect of projectile excitation, we will not
include transfer or inelastic channels of the target. The target
will be regarded as a spherical nucleus with spin zero.
Afterwards, we will estimate the effect of target excitations on
the total fusion cross section. We expect both that such an effect
could be important at energies just around the Coulomb barrier and
that it is similar for the two lithium isotopes.

We would like to stress that in the present calculations the
imaginary parts of the off-diagonal couplings have been neglected,
while the diagonal couplings include imaginary parts. The reason
for not including the imaginary part of the off-diagonal couplings
is that they produce numerical instabilities. The hermiticity of
the symmetric breakup matrix is violated when large values of
off-diagonal imaginary couplings are included. Those imaginary
couplings describe absorption occurring during the transitions
between the channels. Following Ref. \cite{Satchler}, we expect
that these couplings weakly affect the total fusion cross section.
In that reference, it was shown that the imaginary off-diagonal
coupling redistributes, among the elastic and nonelastic channels,
the flux that has already penetrated the Coulomb barrier. The
total fusion cross section, however, remained unchanged.

In addition to the Coulomb interaction, the global Woods-Saxon
parametrization given in Ref. \cite{Broglia} for the Christensen
and Winther potential is used for the real part of the optical
potentials between the projectile fragments ($^{6}$Li=$\alpha
$+$d,$ and $^{7}$Li=$\alpha $+$t$) and the target. Those
potentials are given in Table I. The projectile-target bare
potential for a central collision is calculated by the
single-folding of the projectile fragments-target monopole (real)
interactions with the $^{6,7}$Li ground-state densities defined in
terms of the ground-state wave functions. In the following, by
Coulomb barrier $V_{B}$ we mean the barrier of that potential.

The couplings are taken into account up to a projectile-target
radial distance $R_{\text{coup}}=150$ fm. Partial waves for the
projectile-target relative motion up to only $L_{\max } \sim 30$
(partial-wave total fusion cross section $\lesssim 10^{-3}$ mb)
are included in the calculation.

\begin{table}
\caption{Potentials between the projectile fragments and the
target are shown along with those to describe the projectile
states (g.s.-ground state, res.-resonances, b.s.-bound states).
The potential depths are given in MeV, and the radii and
diffusenesses in fm.}
\begin{ruledtabular}
\begin{tabular}{ccccccc}
Pot.& V$_{0}$  & r$_{0}$   & a$_{0}$   &
V$_{0}^{s.o.}$  & r$_{0}^{s.o.}$  & a$_{0}^{s.o.}$ \\
\hline
$\alpha $-$^{59}$Co & -31.1368 & 1.1273 & 0.63 & - & - & - \\
$\alpha $-$^{209}$Bi & -33.9497 & 1.1675 & 0.63 & - & - & - \\
$d$-$^{59}$Co & -19.9336 & 1.0895 & 0.63 & - & - & - \\
$d$-$^{209}$Bi & -21.0497 & 1.1422 & 0.63 & - & - & - \\
$t$-$^{59}$Co & -25.2677 & 1.1128 & 0.63 & - & - & - \\
$t$-$^{209}$Bi & -25.1141 & 1.1578 & 0.63 & - & - & - \\
$^{6}$Li (g.s.) & -78.46 & 1.15 & 0.7 & - & - & - \\
$^{6}$Li (res.) & -80.0 & 1.15 & 0.7 & 2.5
& 1.15 & 0.7 \\
$^{7}$Li (b.s.) & -108.1 & 1.15 & 0.7 &
0.9875 & 1.15 & 0.7 \\
$^{7}$Li (res.) & -109.89 & 1.15 & 0.7 & 1.6122
& 1.15 & 0.7 \\
\end{tabular}
\end{ruledtabular}
\end{table}

The bound states of the two-body $^{6}$Li ($^{7}$Li) projectile and the
single energy scattering wave functions which form the continuum bins
\cite{Alexis1}, are obtained by solving a Schr\"{o}dinger equation with the $%
\alpha $-$d$ ($\alpha $-$t$) potential $V_{\alpha \text{-}d}^{l}$ ($%
V_{\alpha \text{-}t}^{l}$), which may be $l$ dependent. The continuum states
with a given partial wave $l$ have been consistently generated either by the
same potential as that of the bound state of the same orbital angular
momentum $l$ or by the potential generating the unbound resonances. The
continuum (non-resonant and resonant) breakup subspace is discretized in
equally spaced momentum bins with respect to the momentum $\hbar k$ of the
$\alpha $-$d$ ($\alpha $-$t$) relative motion. The bin widths are suitably
modified in the presence of the resonant states in order to avoid double
counting.

The $J^{\pi }=1^{+}$ ($l=0$ coupled to the spin of the deuteron
$s=1$) ground state of $^{6}$Li with a binding energy of $-1.47$
MeV can be generated by a Woods-Saxon potential given in Table I.
The $d$-state ($l=2$,$s=1$) component \cite{Zhukov} of the g.s.
wave function has been neglected. The $3^{+},2^{+}$ and $1^{+}$
($l=2$ coupled to the spin of the deuteron $s=1$) unbound resonant
states of $^{6}$Li can be obtained with a Woods-Saxon potential
including a spin-orbit term with the same geometry (see Table I).
The energies and widths obtained for those resonances are compared
with the experimental \cite{exp_resonances} values in Table II.

\begin{table}
\caption{Energies $E_{res}$ (MeV) and widths $\Gamma _{res}$ (MeV) of the
calculated resonances are compared with the experimental values
\cite{exp_resonances}.}
\begin{ruledtabular}
\begin{tabular}{cccccc}
Proj.   & Res.   & $E_{res}$   & $\Gamma _{res}$   &
$E_{res}^{\exp.}$   & $\Gamma _{res}^{\exp .}$
\\ \hline
$^{6}$Li   & 3$^{+}$   & 0.73   & 0.034    & 0.716   & 0.024 \\
$^{6}$Li   & 2$^{+}$   & 3.09   & 1.3      & 2.84    & 1.7 \\
$^{6}$Li   & 1$^{+}$   & 4.67   & 4.2      & 4.18    & 1.5 \\
$^{7}$Li   & 7/2$^{-}$ & 2.17   & 0.071    & 2.16    & 0.093 \\
$^{7}$Li   & 5/2$^{-}$ & 4.09   & 0.6      & 4.21    & 0.88 \\
\end{tabular}
\end{ruledtabular}
\end{table}

In the case of $^{7}$Li, the $3/2^{-}$ ground state and the bound $1/2^{-}$
excited state ($l=1$ coupled
to the spin of the triton $s=1/2$) with binding energies of $-2.47$ MeV
and $-1.99$ MeV, respectively, can also be reproduced by a Woods-Saxon
potential with a spin-orbit term (see Table I).
The $7/2^{-}$ and $5/2^{-}$ ($l=3$ coupled to the spin of the triton
$s=1/2$) unbound resonant states are calculated in the same way
(see Table I). The energies and widths obtained for those
resonances are included in Table II.

For the reactions with $^{6}$Li, we obtain converged total fusion
cross sections \cite{Alexis1} using (i) a maximum energy of the
continuum states of $8$ MeV for energies well above the Coulomb
barrier and of $6$ MeV for energies around the barrier, (ii)
continuum partial waves up to $l=2$ waves for a density of the
continuum discretization of $2$ bins/MeV ($l=0,1$); $7.7$ bins/MeV
and $1.92$ bins/MeV below and above the $3^{+}$ resonance,
respectively, $10$ bins/MeV inside the resonance; $2.5$ bins/MeV
and $2$ bins/MeV below and above the $2^{+}$ resonance,
respectively, $2.5$ bins/MeV inside the resonance; for $1^{+}$
continuum states the density of the discretization is the same as
that for $2^{+}$ states, (iii) the projectile fragments-target
potential multipoles up to the quadrupole term ($K\leq 2$).

For $^{7}$Li, converged total fusion cross sections are obtained
using (i) the same cutoff for the maximum energy of the continuum states
as that for $^{6}$Li, (ii) continuum partial waves up to $l=3$ waves for a
density of the
continuum discretization of $2$ bins/MeV ($l=0,1,2$); $7.7$ bins/MeV and $%
1.92$ bins/MeV below and above the $7/2^{-}$ resonance, respectively, $10$
bins/MeV inside the resonance; $2.5$ bins/MeV and $2$ bins/MeV below and
above the $5/2^{-}$ resonance, respectively, $2.5$ bins/MeV inside the
resonance, (iii) the projectile fragments-target potential multipoles up to
the octupole term ($K\leq 3$).

\section{Results and discussion}

\subsection{$^{6,7}$Li+$^{59}$Co}

Fig.1 is used as an example to show the convergence of the total
fusion cross sections of $^{6}$Li+$^{59}$Co with respect to the
number $l$ of partial waves in the continuum along with potential
multipoles $K$. The maximum energy E$_{max}$ of the continuum
states and the density of the continuum discretization are as
mentioned above. In particular, we would like to point out that
converged results at energies just around the Coulomb barrier are
obtained with a small number of continuum partial waves, i.e. up
to $d$-waves in the present example, in contrast to that, it was
claimed in Ref. \cite{Yabana}. In the following, total fusion
cross sections refer to converged values.

\begin{figure}
\includegraphics[width=8.66cm]{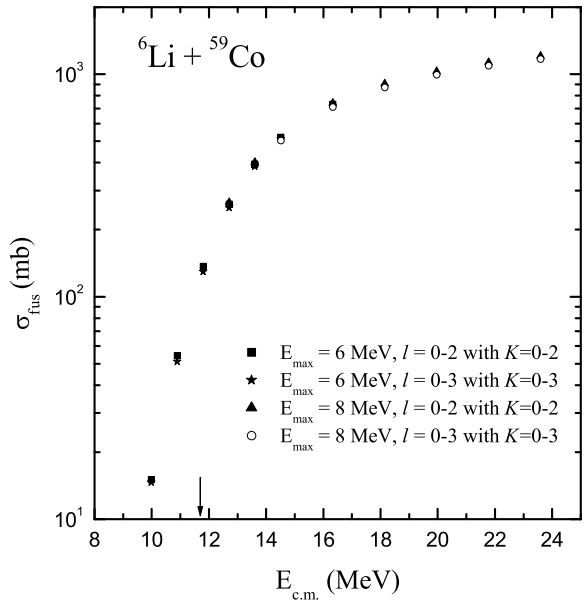}
\caption{Convergence of the total fusion cross sections for
$^{6}$Li+$^{59}$Co with regard to the number $l$ of continuum
partial waves along with the potential multipoles $K$. The arrow
indicates the value of the Coulomb barrier. See text for further
details.}
\end{figure}

 Fig. 2 shows total fusion excitation functions for
$^{6}$Li+$^{59}$Co (full dots with solid curve) and
$^{7}$Li+$^{59}$Co (full triangles with dashed curve), which are
normalized with the corresponding cross sections in absence of
couplings to breakup channels. For each reaction, the incident
energy is normalized with its Coulomb barrier $V_B$. The Coulomb
barrier for $^{6}$Li+$^{59}$Co is $V_B=11.74$ MeV, whereas for
$^{7}$Li+$^{59}$Co it is $V_B=11.68$ MeV. These barriers are
similar to those ($11.5$ MeV and $11.35$ MeV, respectively)
calculated using the double-folding procedure with the Skyrme-type
nucleon-nucleon interaction \cite{DF}. The total fusion excitation
functions without breakup include all reorientation couplings of
$^{6,7}$Li. In case of $^{7}$Li, it also includes the coupling to
the bound $1/2^{-}$ first excited state. Since $^{6}$Li is a
spherical nucleus in its ground state and, therefore, the
quadrupole term of the projectile fragments-target potentials is
zero, no reorientation effects occur for $^{6}$Li. The total
fusion excitation functions with breakup include all continuum and
reorientation couplings.

\begin{figure}
\includegraphics[width=8.66cm]{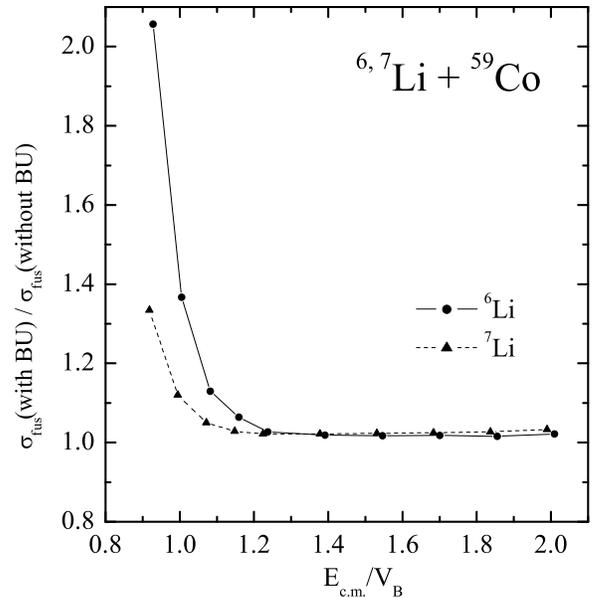}
\caption{Total fusion excitation functions for $^{6}$Li+$^{59}$Co
(full dots) and $^{7}$Li+$^{59}$Co (full triangles), which are
normalized with the cross sections in absence of couplings to
breakup (BU) channels. For each reaction, the incident energy is
normalized with the Coulomb barrier $V_B$ of the bare potential.
The calculated values are connected with curves to guide the eye.
See text for further details.}
\end{figure}

The breakup enhances the total fusion cross section at energies
just around the barrier, whereas it hardly affects (an enhancement
by $\sim 2\%$) the total fusion at energies well above the
barrier. The fusion enhancement around the barrier is larger for
$^{6}$Li than for $^{7}$Li, and it correlates with the smaller
$\alpha$-breakup threshold for $^{6}$Li. Here, the enhancement
factor strongly depends on the incident energy. This enhancement
is caused by the bound-continuum couplings \cite{Alexis1} which
dominate the suppression effect of the continuum-continuum
couplings.

\begin{figure}
\includegraphics[width=8.66cm]{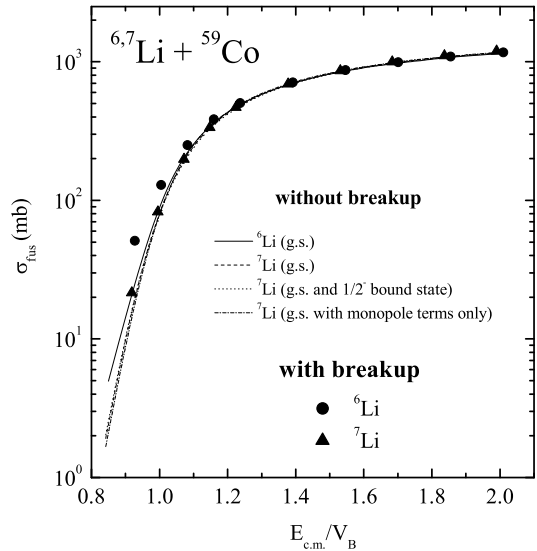}
\caption{Total fusion excitation functions for $^{6}$Li+$^{59}$Co
are compared to those for $^{7}$Li+$^{59}$Co. For each reaction,
the incident energy is normalized with the Coulomb barrier of the
bare potential. (fusion without breakup) The reorientation
couplings are only included. (fusion with breakup) All continuum
and reorientation couplings are included. See text for further
details.}
\end{figure}

In Fig. 3, we compare the total fusion excitation functions of the
two reactions. For each reaction, the incident energy is also
normalized with the Coulomb barrier $V_B$ of the bare potential.
We only include the reorientation couplings in fusion without
breakup. All continuum and reorientation couplings are included in
fusion with breakup. We can observe that the total fusion
excitation functions without breakup are practically the same for
$^{6,7}$Li. The mass difference between $^{6,7}$Li explains the
remaining difference between their fusion excitation functions at
energies well below the barrier. Both the $^{7}$Li ground-state
reorientation effect (comparing the dashed curve to the
dash-dotted one) and the effect of the coupling to its bound
$1/2^{-}$ excited state (comparing the dashed curve to the dotted
one) on total fusion are very weak.

In Fig. 3, we can also observe that the inclusion of the couplings
to the breakup channels notably increases the difference between
the total fusion excitation functions of the two systems at
energies just around the barrier. However, at energies well above
the barrier the total fusion cross sections are practically the
same. We conclude that the breakup is the main reason for the
difference between the total fusion cross sections of $^{6,7}$Li
on $^{59}$Co.

A crude estimation of the effect of $^{59}$Co excitations on the
total fusion cross section was done by (i) fitting the converged
total fusion cross sections of Fig. 3 (lower part) in a single
(elastic) channel calculation by finding an appropriate
projectile-target real Woods-Saxon potential with an energy
dependent depth and the geometry $r_0=1.179$ fm and $a=0.658$ fm,
and then (ii) including the target excitations as in Refs.
\cite{FRESCO,target_exc}. We include the couplings to the first
three levels of the ground-state rotational band (Table III) of
$^{59}$Co \cite{Baglin}. This estimation reveals that the effect
is very weak and similar for the two lithium isotopes. Total
fusion cross sections are increased by $\sim 5\%$ for energies
around the barrier, while they remain constant for energies well
above the barrier.

\begin{figure}
\includegraphics[width=8.66cm]{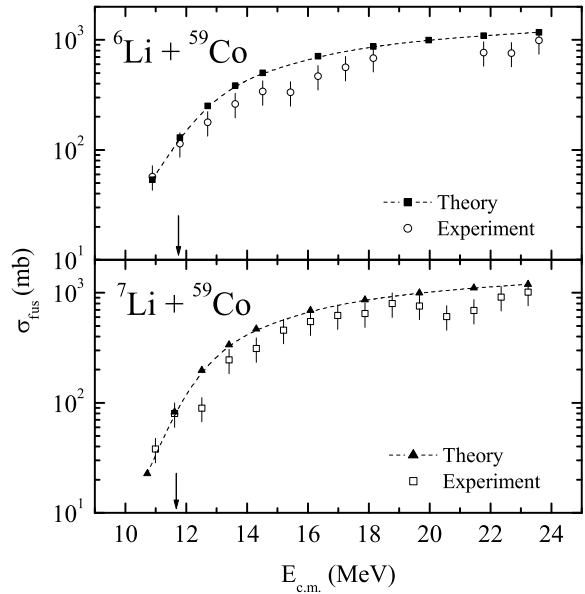}
\caption{Calculated (full squares and triangles) total fusion
excitation functions for $^{6,7}$Li+$^{59}$Co are compared with
the experimental data \cite{Beck_new} (open circles and squares).
The calculated values are connected with curves to guide the eye.
The arrows indicate the Coulomb barriers of the bare potentials.
See text for further details.}
\end{figure}

In Fig. 4, the calculated (full squares and triangles) total
fusion cross sections of Fig. 3 (lower part) including the effect
of $^{59}$Co excitations are compared to the experimental data
\cite{Beck_new} (open circles and squares). The agreement is good
at energies just around the barrier (arrows), but a slight
overestimation is observed at energies well above the barrier.
However, the ratio of the theoretical cross sections of the two
systems agrees very well with the ratio of their experimental
cross sections (see Fig. 3 of \cite{Beck_new}) at all energies.

\begin{table}
\caption{Experimental values for (i) the first three levels of the
ground-state rotational band of the $^{59}$Co target
\cite{Baglin}, and for (ii) the excitations of $^{209}$Bi included
in the calculation \cite{Martin}.}
\begin{ruledtabular}
\begin{tabular}{cccccc}
Target & Energy (MeV) & $I^{\pi}$ & B(E$\lambda$, I $\to $ g.s.) (W.u.) \\
\hline
$^{59}$Co & 0.0 & 7/2$^{-}$ & - \\
          & 1.19 & 9/2$^{-}$ &  13.0 (E2)\\
          & 1.46 & 11/2$^{-}$ &  5.4 (E2)\\
$^{209}$Bi & 0.0 & 9/2$^{-}$ & - \\
           & 2.493 & 3/2$^{+}$ & 16.0 (E3) \\
           & 2.564 & 9/2$^{+}$ & 28.0 (E3) \\
           & 2.583 & 7/2$^{+}$ & 25.0 (E3) \\
           & 2.599 & 11/2$^{+}$ & 30.0 (E3) \\
           & 2.6 & 13/2$^{+}$ & 22.0 (E3) \\
           & 2.617 & 5/2$^{+}$ & 22.0 (E3) \\
           & 2.741 & 15/2$^{+}$ & 25.0 (E3) \\
\end{tabular}
\end{ruledtabular}
\end{table}

\subsection{$^{6,7}$Li+$^{209}$Bi}

Fig. 5 shows, like Fig. 2, total fusion excitation functions for
$^{6}$Li+$^{209}$Bi (full dots with solid curve) and
$^{7}$Li+$^{209}$Bi (full triangles with dashed curve), which are
normalized with the corresponding cross sections without breakup.
The incident energies are also normalized with the barrier
($V_B=29.71$ MeV for $^{6}$Li, and $V_B=29.57$ MeV for $^{7}$Li)
of the calculated bare potentials. These barriers are similar to
those measured in Ref. \cite{Dasgupta_new} ($V_B=30.1 \pm 0.3$ MeV
for $^{6}$Li, and $V_B=29.7 \pm 0.2$ MeV for $^{7}$Li) and to
those calculated \cite{DF} ($V_B=29.8$ MeV for $^{6}$Li, and
$V_B=29.5$ MeV for $^{7}$Li) with the double-folding procedure.

Fusion enhancement occurs at energies just around the barrier,
while breakup has very little effect (an enhancement by $\sim
3.5\%$) on total fusion at energies well above the barrier. The
difference between the two $\alpha$-breakup thresholds for
$^{6,7}$Li is also revealed in the value of their enhancement
factors around the barrier. These factors also depend strongly on
the decreasing incident energy below the barrier. Comparing this
figure with Fig. 2, we can observe that the breakup effect on
total fusion is stronger for the $^{209}$Bi target than for
$^{59}$Co, as expected.

\begin{figure}
\includegraphics[width=8.66cm]{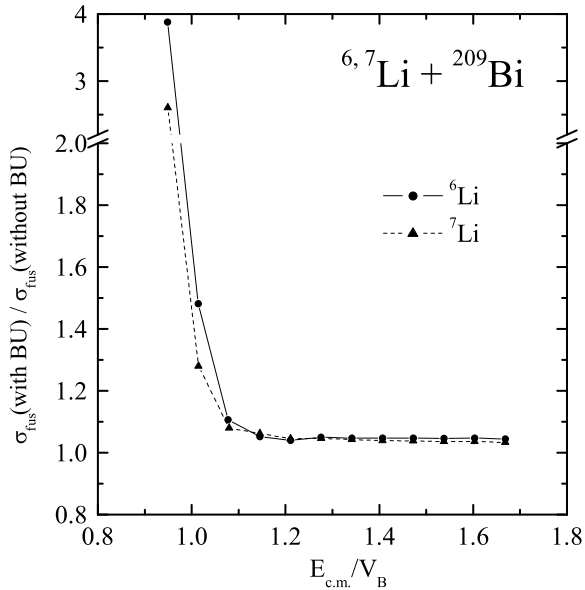}
\caption{The same as Fig. 2, but for $^{6}$Li+$^{209}$Bi (full
dots) and $^{7}$Li+$^{209}$Bi (full triangles). See text for
further details.}
\end{figure}

In Fig. 6, the total fusion excitation functions of the two
reactions are compared to each other (as in Fig. 3). It is
observed that the differences between the total fusion excitation
functions, caused by the bound-state structure effects of
$^{6,7}$Li, are very small and similar to those with the $^{59}$Co
target. The effect of the $^{7}$Li ground-state reorientation
couplings (comparing the dashed curve to the dash-dotted one) and
of the coupling to its bound $1/2^{-}$ excited state (comparing
the dashed curve to the dotted one) on total fusion are also very
weak.

\begin{figure}
\includegraphics[width=8.66cm]{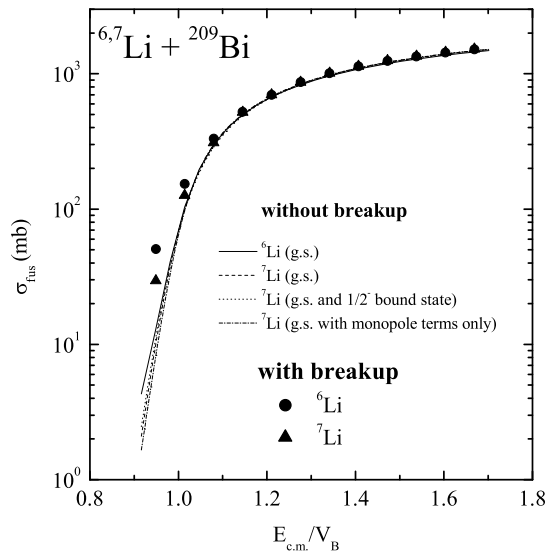}
\caption{The same as Fig. 3, but for $^{6,7}$Li+$^{209}$Bi. See
text for further details.}
\end{figure}

In Fig. 6, it is also shown that the breakup increases the
difference between the total fusion cross sections for $^{6,7}$Li
at energies just around the barrier. At energies well above the
barrier, the two systems show very similar total fusion excitation
functions. Fig. 6 also indicates, like Fig. 3, that the breakup
causes the difference between the total fusion cross sections of
$^{6,7}$Li.

\begin{figure}
\includegraphics[width=8.66cm]{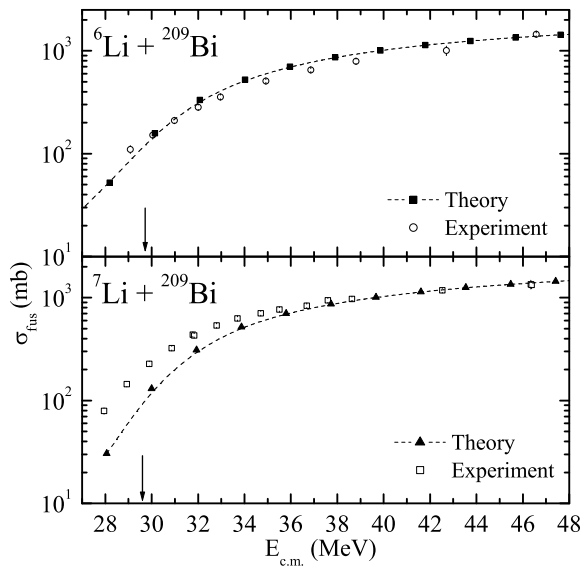}
\caption{The same as Fig. 4, but for $^{6,7}$Li+$^{209}$Bi. The
experimental data are from Ref. \cite{Dasgupta_new}. See text for
further details.}
\end{figure}

We also estimated the effect of target excitations on total
fusion. It was done in the same way as for $^{59}$Co. Since the
couplings to the first two excited states of single-particle
structure is rather weak \cite{Martin}, we only include couplings
to the collective multiplet
[$^{208}$Pb(3$^{-}$)$\otimes$1h$_{9/2}$]$_{J^{\pi}}$ (Table III)
with energies ranging from 2.493 to 2.741 MeV \cite{Martin}.
Moreover, we assume collective transitions to these excited states
due to their complexity (combined collective and single-particle
dynamics). Total fusion cross sections remain the same for
energies well above the barrier, whereas they are increased by
$\sim 3\%$ around the barrier for both $^{6}$Li and $^{7}$Li.

Fig. 7 shows a comparison between the calculated (full squares and
triangles) total fusion excitation functions including the effect
of target excitations and the experimental data
\cite{Dasgupta_new} (open circles and squares). The experimental
data for $^{6}$Li (upper part, open circles) are well reproduced,
but it is not the case for $^{7}$Li (lower part, open squares).
Theoretical results for $^{7}$Li underestimate the experimental
values at energies around the barrier, but the agreement is good
at energies well above the barrier.

In contrast to $^{6,7}$Li+$^{59}$Co \cite{Beck_new}, experimental
total fusion cross sections for $^{7}$Li+$^{209}$Bi are larger
than those for $^{6}$Li+$^{209}$Bi around the barrier. The
experiment \cite{Dasgupta_new} shows that it is because the direct
production of $^{210,211}$Po evaporation residue (contributing to
the incomplete fusion yield) is notably larger with $^{7}$Li than
with $^{6}$Li. These evaporation residues can be produced by the
capture of (i) the deuteron or proton (proton capture following
dissociation of $d \to p+n$) for $^{6}$Li, and of (ii) the triton
for $^{7}$Li. A stripping breakup process \cite{Stripping} does
not seem to explain the larger production of $^{210,211}$Po with
$^{7}$Li, because the triton binding energy in $^{7}$Li is larger
than the deuteron one in $^{6}$Li. It would be interesting to
measure deuteron and triton transfer cross sections for these
reactions to clarify the reason for the difference of their
$^{210,211}$Po yields. We would like to stress that couplings to
these transfer channels were not included in the present
calculations.

\subsection{Effect of the off-diagonal imaginary couplings;
comparison to other approaches}

As an example, Fig. 8 shows the effect of the off-diagonal
imaginary couplings on the total fusion of $^{6}$Li+$^{59}$Co.
This is the only system (the lightest one studied) for which we
have obtained stable results when all imaginary couplings are
included, as the off-diagonal imaginary
couplings are not too strong in this case. In Fig. 8, we also compare the
total fusion excitation function obtained with the present
approach (using two optical potentials) to that obtained with our
previous method in Ref. \cite{Alexis1} (using one imaginary
potential defined in the c.m. coordinate of $^{6}$Li). The cross
sections neglecting the off-diagonal imaginary couplings (full
squares in Fig. 8) are the same as shown with full dots in Fig. 3.
The cross sections including all imaginary couplings are presented
with full stars. The off-diagonal imaginary couplings slightly
reduce the total fusion cross sections (by $\sim 13\%$ around the
barrier and by $\sim 6\%$ well above the barrier). However, the
cross sections obtained with the present approach (full squares)
differ considerably from those (full triangles) calculated with
the method from Ref. \cite{Alexis1}, as expected for the $^{6}$Li
projectile (see introduction). The results indicate that there are
many events where one of the fragments of $^{6}$Li is captured,
but the c.m. of the projectile does not reach the absorption
(fusion) region. The imaginary potential used in the method from
Ref. \cite{Alexis1} has the same shape and magnitude as the
imaginary part of the optical potentials in the present approach.

\begin{figure}
\includegraphics[width=8.66cm]{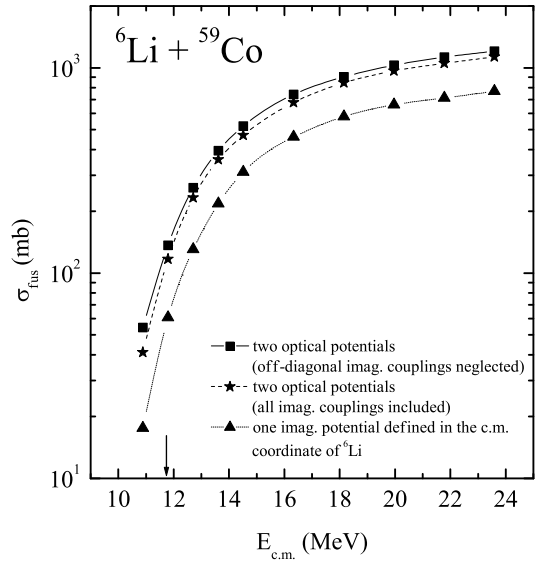}
\caption{Total fusion excitation functions for $^{6}$Li+$^{59}$Co
using different approaches. The calculated values are connected
with curves to guide the eye. The arrow indicates the value of the
Coulomb barrier. See text for further details.}
\end{figure}

\begin{figure}
\includegraphics[width=8.66cm]{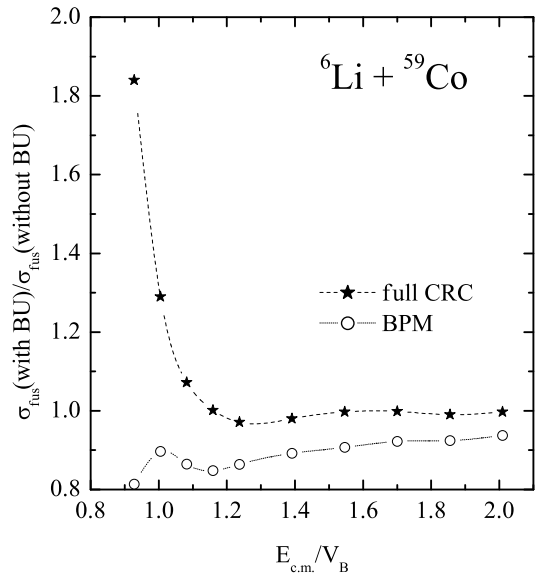}
\caption{ The same as Fig. 2 and 5, but comparing total fusion
excitation functions for $^{6}$Li+$^{59}$Co using two different
approaches, namely full coupled channels calculations (CRC) and
the barrier penetration model (BPM). The calculated values are
connected with curves to guide the eye. See text for further
details.}
\end{figure}

In Fig 9, we compare the total fusion excitation function (open
circles) obtained with the barrier penetration model (BPM)
\cite{Satchler} to that obtained with full coupled channels
calculations when all imaginary couplings are included (full stars
in Fig. 8). The total fusion excitation functions are normalized
with the cross sections in absence of couplings to breakup
channels. The BPM model assumes that all the flux that penetrates
a single barrier (defined for these calculations as the sum of the
bare potential and the real part of a local dynamic polarization
potential (DPP) \cite{target_exc}) leads to fusion. The DPP
potential includes the effect of couplings to breakup channels and
was extracted from the CDCC calculation shown with full stars in
Fig. 8. This approach to fusion has been extensively used in the
last few years \cite{Keeley1,Keeley2} to predict total fusion
cross sections with weakly bound projectiles, e.g. very recently
for $^{6}$Li and $^{6}$He on the $^{208}$Pb target \cite{Keeley2}.
It is observed that the BPM cross sections (open circles)
underestimate (by $\sim 40\%$ just around the barrier and by $\sim
8\%$ well above the barrier) the actual cross sections (full
stars), which is particularly relevant at energies just around the
barrier. This is because the BPM approach does not include fusion
within the projectile-target barrier \cite{target_exc}. In
contrast to the full coupled channel calculations (full stars),
the BPM approach (open circles) shows total fusion suppression for
the whole range of incident energies studied (by $\sim 14\%$ just
around the barrier and by $\sim 9\%$ well above the barrier). The
present results shows that the BPM approach is particularly
inappropriate at energies just around the barrier and also yields
a rather inaccurate answer to the question how breakup affects
total fusion.

With the above comparisons, it is important to note that to
realistically predict total fusion cross sections the fusion model
should be carefully chosen depending on the cluster structure of
the weakly bound projectile and on what the experiment is expected
to measure (in order to ensure a realistic comparison of
theoretical and experimental data).

\section{Summary and conclusions}

Total (complete + incomplete) fusion excitation functions of
$^{6,7}$Li on $^{59}$Co and $^{209}$Bi targets at Coulomb barrier
energies are obtained using full coupled channels calculations
with a new CDCC method of calculating fusion, that has
short-range fusion potentials for each fragment separately. The
realistic prediction of total fusion cross sections requires the
selection of an appropriate fusion model which depends on both the
cluster structure of the weakly bound projectile and what the
experiment is supposed to measure.

The effect of breakup on fusion can be observed in the total
fusion excitation function. The breakup enhances the total fusion
of $^{6,7}$Li at energies just around the barrier, whereas it has
very little effect on total fusion at energies well above the
barrier. The fusion enhancement factor strongly depends on the
decreasing incident energy below the barrier. The fusion
enhancement is larger for the reaction with $^{6}$Li than that
with $^{7}$Li, and it is correlated with the smaller
$\alpha$-breakup threshold of $^{6}$Li. The difference between the
bound-state structures of $^{6,7}$Li does not produce large
difference between their total fusion excitation functions. The
effect of the $^{7}$Li ground-state reorientation couplings on
total fusion is very weak. A crude estimation reveals that the
effect of target excitations on total fusion is weak and similar
for $^{6,7}$Li. The experimental data for $^{6,7}$Li+$^{59}$Co as
well as for $^{6}$Li+$^{209}$Bi are well reproduced. The breakup
notably causes the difference between the experimental total
fusion cross sections of $^{6,7}$Li on $^{59}$Co, but it is not
the case for the $^{209}$Bi target. Experiments focused on the
deuteron and triton transfer cross sections are important to
understand the difference between the total fusion of $^{6,7}$Li
on $^{209}$Bi. Work is in progress to study $^{6}$He induced
fusion reactions since low-energy radioactive beams have become
available for new experiments in selected facilities
\cite{Kolata,Rehm,Trota,Signorini}.

\begin{center}
{\bf AKNOWLEDGEMENTS}
\end{center}

The authors thank M. Dasgupta for providing experimental data in
tabulated form. We would also like to acknowledge G.G. Adamian,
N.V. Antonenko and S.J. Sanders for a careful reading of the
manuscript.

\end{document}